\documentclass[conference]{IEEEtran}

\usepackage{cite}
\usepackage{graphicx}
\usepackage{color}
\usepackage{caption}
\usepackage{subcaption}
\usepackage{epsfig}
\usepackage{epstopdf}
\usepackage{float}
\usepackage{multicol}
\usepackage{array} 
\usepackage{paralist} 
\usepackage{verbatim}
\usepackage[cmex10]{amsmath}
\usepackage{amssymb}
\usepackage{algorithm}
\usepackage{algorithmic}
\usepackage{mdwmath}
\usepackage{mdwtab}
\usepackage{fixltx2e}
\usepackage{stfloats}
\usepackage[section]{placeins}
\normalsize
\begin{document}
%
\title{A Simple Angle of Arrival Estimation Scheme}


%
\author{\IEEEauthorblockN{Ahmed Badawy\IEEEauthorrefmark{1}\IEEEauthorrefmark{2},
Tamer Khattab\IEEEauthorrefmark{2},
Daniele Trinchero\IEEEauthorrefmark{1},
Tarek ElFouly\IEEEauthorrefmark{3}, and
Amr Mohamed\IEEEauthorrefmark{3},
}
\IEEEauthorblockA{\IEEEauthorrefmark{1}Politecnico di Torino, DET - iXem Lab. (ahmed.badawy,daniele.trinchero@polito.it)}
\IEEEauthorblockA{\IEEEauthorrefmark{2}Qatar University, Electrical Engineering Dept.
(tkhattab@qu.edu.qa)}
\IEEEauthorblockA{\IEEEauthorrefmark{3}Qatar University, Computer Engineering Dept. (tarekfouly,amrm@qu.edu.qa)}
}
\maketitle
\begin{abstract}
 We propose an intuitive, simple and hardware friendly, yet surprisingly novel and efficient, received signal's angle of arrival (AoA) estimation scheme.  Our intuitive, two-phases cross-correlation based scheme relies on a switched beam antenna array, which is used to collect an omni-directional signal using few elements of the antenna array in the first phase. In the second phase, the scheme switches the main beam of the antenna array to scan the angular region of interest. The collected signal from each beam (direction or angle) is cross correlated with the omni-directional signal. The cross-correlation coefficient will be the highest at the correct AoA and relatively negligible elsewhere.  The proposed scheme simplicity stems from its low computational complexity (only cross-correlation and comparison operations are required) and its independence of the transmitted signal structure (does not require information about the transmitted signal).  The proposed scheme requires a receiver with switched beam antenna array, which can be attached to a single radio frequency chain through phase shifters, hence, its hardware friendliness.  The high efficiency of our system can be observed by comparing its performance with the literature's best performing MUSIC algorithm.  The comparison demonstrates that our scheme outperforms the MUSIC algorithm, specially at low SNR levels. Moreover, the number of sources that can be detected using our scheme is bound by the number of switched beams, rather than the number of antenna elements in the case of the MUSIC algorithm.
\end{abstract}
\begin{IEEEkeywords} angle of arrival Estimation, AoA, Switched Beam, direction of arrival estimation, DoA.\end{IEEEkeywords}
\section{Introduction}

AoA estimation is a process that determines the direction of arrival of a received signal by processing the signal impinging on an antenna array. Applications of estimating the AoA include beamforming, tracking and localization. The subject of AoA has been extensively studied in the literature \cite{AoA_book_Tuncer, AoA_book_Balanis, AoA_book_Chen, AoA_book_chandran, AoA_book_foutz}. Referring to the aforementioned books; from a system perspective, one can categorize AoA estimation systems into two main categories: \begin{enumerate}
              \item \textbf{Switched beam system (SBS)} which uses a fixed number of beams to scan the azimuth plane. The AoA is the angle of the beam with the highest received power or signal strength (RSS). The SBS is easy to implement since it requires a single receiver and no baseband signal processing technique is needed to estimate the AoA. In other words, the hardware and computational complexities of SBSs are low. On the contrary, if the power of the received signal is lower than the receiver sensitivity, i.e., at low signal to noise ratio (SNR), SBS will fail to estimate the AoA.
              \item \textbf{Adaptive array system (AAS)} can steer the beam in any desired direction by setting the weights across the antenna array elements. AAS requires $M$ receivers to estimate the AoA, where $M$ is the number of antenna elements in the array. The AoA estimation techniques that use AAS can operate at lower SNRs than the SBS technique but has higher hardware and computational complexities. The AoA estimation techniques that use AAS can be divided into two main groups: \begin{enumerate}
                  \item \textbf{Classical AoA techniques} which include two main techniques which are Delay and Sum, also known as Bartlett \cite{BARTLETT} and Minimum Variance Distortionless (MVDR), also known as Capon \cite{Capon}. By steering the beams electronically and estimating the power spectrum of the received signal, the AoAs are estimated as being the peaks in the spatial power spectrum. The main drawback of the Bartlett technique is that signal impinging with angular separation less than $2\pi/M$ can not be resolved. The Capon technique relatively solves the angular resolution drawback of the Bartlett method on the cost of more baseband processing needed for matrix inversion.
                  \item \textbf{Subspace techniques} which are based on the idea that the signal subspace is orthogonal to the noise subspace. The most widely used technique in this group is the MUltiple SIgnal Classification (MUSIC) \cite{Phd1981}. The MUSIC technique provides the highest angular resolution and can operate at low SNR levels. This comes on the cost that it requires a full a priori knowledge of number of sources and the array response, whether it is measured and stored or computed analytically. The signal and noise subspaces are distinguished through an eigen decomposition operation applied at the covariance matrix of the received signal. This operation requires a substantial computational complexity. 
                  \end{enumerate}
            \end{enumerate}

An example of integrating SBS with other theories to estimate the AoA is presented in \cite{Gotsis09}. Their methodology is based on neural network, in which they transform the AoA problem into a mapping problem. This mapping process is a burden to the processor as well as the memory of the system. Another drawback of their system is that they require a prior knowledge of the number of sources as well as the multiple access scheme adopted between them. They also assumed that a power control scheme is implemented such that the source powers are equal. Such requirements and assumptions limits the deployment of their system to very few scenarios. Exploiting the power ratio between the adjacent beams to estimate the AoA is presented in \cite{Ozaki09}. Another table driven SBS system is presented in \cite{Lee05}. Both of these techniques do not tackle the drawbacks of the conventional SBS, rather they make its implementation easier.

The concept of using the cross-correlation function to extract features of the signal can be found in many applications; one of which is the passive radar systems. Passive radar systems exploit the transmitters of opportunity such as television signals to detect an airborne target. In passive radar systems, a cross correlation between a reference signal from the first receiver and a directional signal from the second is applied to estimate bistatic range and doppler shift of the target. The AoA has to be estimated before that at the second receiver to place a null in the direction of the reference signal \cite{Ringer99, Berger10} such that the received directional signal is the reflection from the airborne target. Another application is localization, in which a cross correlation between two signals received from two antennas is applied to estimate the time difference of arrival \cite{Chih-Kang92,Hongcheng10,Pengfei09}.

Our contributions in this work as compared to available literature are as follows: We propose a new AoA estimation system that is based on beam switching. Our system cross correlates an omni-directional signal with narrow beam signals to estimate the AoA. The cross correlation between the omni-directional signal and the signals received from the switched beams is the highest at the true AoA and relatively negligible otherwise. We compare the proposed mechanism with the MUSIC algorithm and show the at our proposed solution outperforms it, particularly for low SNRs. We also compare, qualitatively, the complexity of our approach with the existing one and conclude that our approach again has lower hardware and computational complexity. To the best of the authors' knowledge using the cross correlation coefficient between an omni-directional signal and directed beam signal to estimate the AoA has not been presented in the literature before.

The rest of this paper is organized as follows: In Section II our system model is presented. An overview of the existing AoA estimation techniques is presented in section III.  We then propose our cross correlation based AoA estimation system in Section IV. The paper is then concluded in section VII.

\section{System Model}
\begin{figure}
\centering
\includegraphics[width=2.5in]{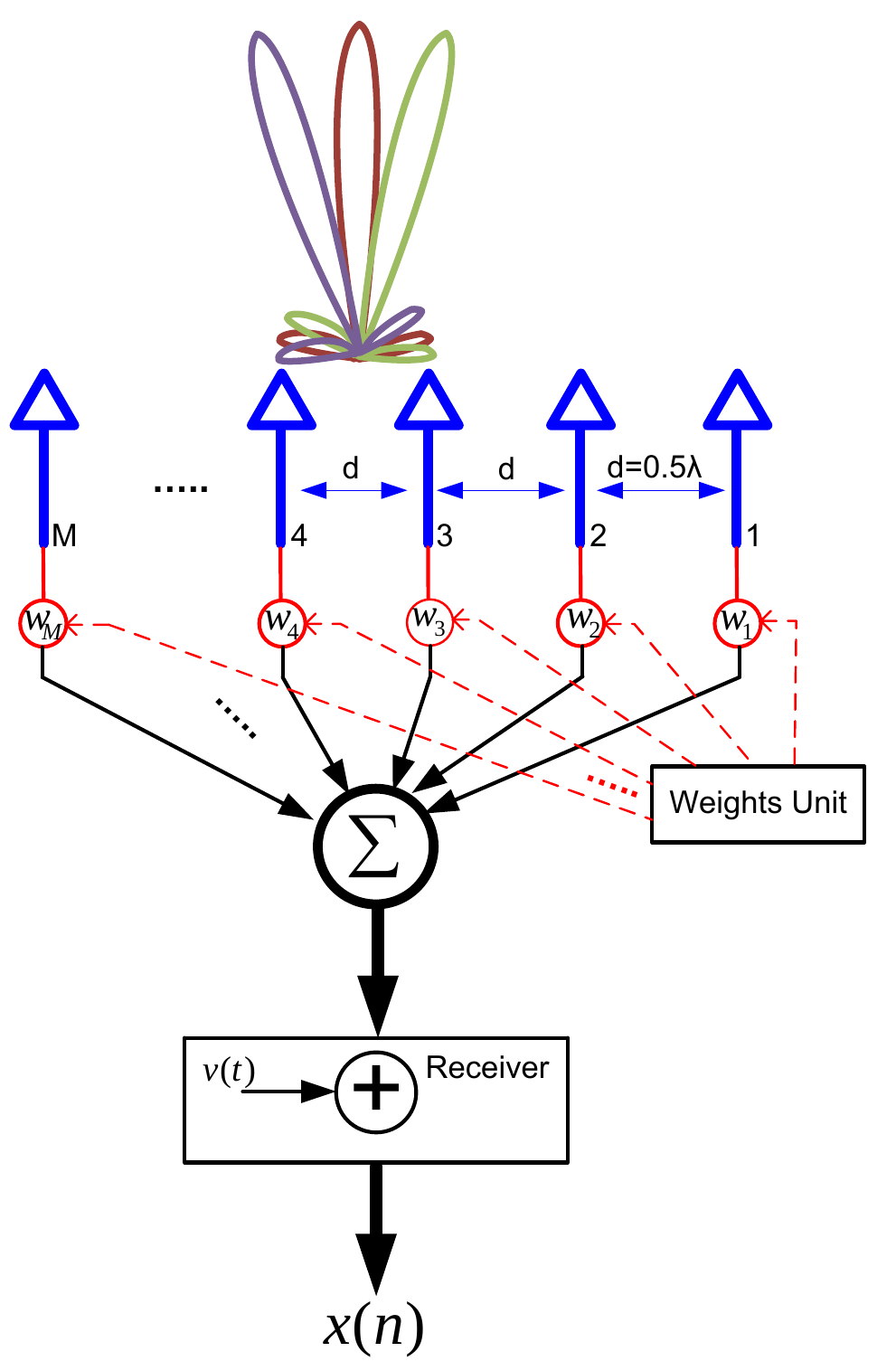}
\caption{Our SBS for $M$ antenna elements.}
\label{fig_000}
\end{figure}
In our system model, we assume that there exists a source that transmits signal $s(t)$. Our receiver is equipped with an SBS presented in Figure \ref{fig_000} consisting of $M$ antenna elements, separated by a fixed separation $d$ and operating at frequency $f$. Our antenna array has an array response vector also known as steering vector $a(\phi)\in \mathbb{C}^{M }$, where $\phi$ is the azimuth angle. The received signal $x(t)$
is then:
\begin{align}
  x_k(t)=\sum_{M}{w_k s(t)}+v(t)
\end{align}

where $v(t)$ is the additive white Gaussian noise (AWGN), $w_k$ is the vector with dimensions $M \times 1$, which is the weight set applied across the antenna array elements such that the steering vector $a(\phi)$ is pointing to an azimuth angle $\phi_k$. The weights are updated to change the direction of the beam, i.e. $\phi_k$, to scan the the angular space of interest, where $k \in [1:K]$, where $K$ is the total number of generated beams. Based on the antenna array formation, whether it is a linear, circular or planar, $a(\phi)$ changes accordingly.
\begin{align}
w=\frac{1}{M}{a(\phi)}
\end{align}
The steering vector for the linear, circular and planar array can be estimated analytically. It's worth noting that our proposed system is independent of the antenna array formation. For a uniform linear array with uniform excitation, $a(\phi)$ can be given by:

\begin{align}
a(\phi)=\left[1, e^{\beta d cos(\phi)}, e^{\beta 2d cos(\phi)},..., e^{\beta (M-1)d cos(\phi)}\right],
\end{align}

where $\beta=\frac{2\pi}{\lambda}$ is the wave number, $\lambda$ is the wavelength and $\phi$ ranges between $[0:\pi]$.
For a uniform circular array (UCA), $a(\phi)$, can be given by \cite{Ioannides05}:

\begin{eqnarray}
a(\phi)&=& [e^{\beta r  \cos (\phi-\phi_1)}, e^{\beta r \cos (\phi-\phi_2)},\\
&&..., e^{\beta r \cos (\phi-\phi_M)} ],
\end{eqnarray}

where
\begin{align}
\phi_m=\frac{2\pi m}{M}, \hspace{0.25cm} m=1,2,..,M
\end{align}
and $\phi$ ranges between $[0:2\pi]$ and $r$ is the radius of the antenna array. The elevation angle is assumed to be 90 degrees in the 1-D angle of arrival estimation techniques. For linear array with uniform excitation, the total number of orthogonal beam that can be generated is $M$. Using the non-uniform excitation such as Dolph-Chebyshev or Taylor \cite{AoA_book_Orfandis,AoA_book_balanis_theory}, it is possible to generate more orthogonal beams for the same number of antenna elements $M$, as we will show later. The digitized version of the received signal is then $x(n)$ for $n=1:N$, where $N$ is the total number of samples, also known as the snapshots.

\section{Existing AoA Estimation Techniques}

As explained in the Introduction, the AAS requires $M$ receivers to down convert the received signals from the $M$ antenna elements to the baseband in order to estimate the AoA. The most popular AAS AoA techniques are the Bartlett, Capon and MUSIC. In the Bartlett technique, the beam is formed across the angular region of interest by applying the weights that correspond to that direction and whichever angle that provides the highest power is the estimated AoA.

The spatial power spectrum $P(\phi)$ for the Bartlett technique is given by \cite{BARTLETT}:
\begin{align}
  P_{Bartlett}(\phi)= a^H(\phi) R_{xx} a(\phi),
\end{align}
where $H$ denotes the Hermitian matrix operation. , $R_{xx}$ is the received signal autocovariance matrix with dimension $M \times M$.  $R_{xx}$ is estimated as:
\begin{align}
R_{xx}=\frac{1}{N}\sum_{n=1}^{N}{x(n)x^H(n)} \label{eqn10}
\end{align}
The Bartlett technique suffers from a low resolution. Capon's technique attempts to overcome the Bartlett low resolution drawback by setting a constraint on the beam former gain to unity in that direction and minimizing the output power from signals coming from all other directions.

The spatial power spectrum for the Capon technique is given by \cite{Capon}:
\begin{align}
  P_{Capon}(\phi)= \frac{1}{a^H(\phi) R_{xx} ^{-1} a(\phi)}
\end{align}

On the other hand, the MUSIC algorithm exploits the orthogonality of the signal and noise subspaces. After an eigenvalue decomposition (EVD) on $R_{xx}$, it can be written as:
\begin{eqnarray}
R_{xx}&=&a R_{ss} a^H + \sigma^2 I\\
&=& U_s \Lambda_s U_s^H +  U_v \Lambda_v U_v^H,
\end{eqnarray}

where $\sigma^2$ is the noise variance, $I$ is the unitary matrix, $U_s$ and $U_v$ are the signal and noise subspaces unitary matrices and $\Lambda_s$ and $\Lambda_v$ are diagonal matrices of the eigenvalues of the signal and noise.

The spatial power spectrum for the MUSIC technique is given by \cite{Phd1981,Phd2}:
\begin{align}
  P_{MUSIC}(\phi)= \frac{1}{a^H(\phi) {P_v} a(\phi)},
\end{align}

where $P_v = U_v U_v^H$.

\begin{figure}
\centering
\includegraphics[width=3.5in]{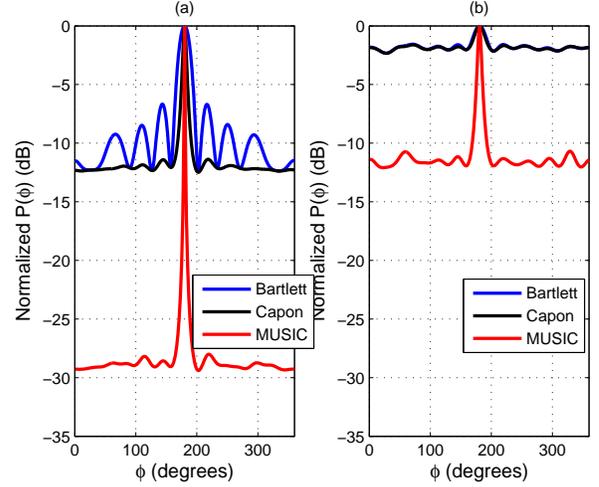}
\caption{Comparison of existing AoA estimation techniques using a uniform circular array for a single source at $\phi = 180$ with $M=16$ and $N=1000$ samples: (a) at SNR = 0 dB and (b) at SNR = -15 dB }
\label{fig_1}
\end{figure}

Figure \ref{fig_1} shows the simulation results for the three algorithms for $M=16$, $N=1000$ for a UCA : at (a) SNR = 0 dB and (b) SNR = -15 dB. It is shown that the MUSIC technique outperforms the two other techniques; achieving a high peak to floor ratio (PFR) of the normalized spatial power of almost 28 dB for the MUSIC, -13 for the Capon and -10 for the Bartlett at SNR = 0 dB. At low SNR, the Capon and Bartlett algorithms almost fail to estimate the AoA achieving PFR of 2 dB while the MUSIC is achieving 12 dB. In other words, the MUSIC can operate at low SNR levels while the Bartlett and Capon will fail to do so.

\begin{figure}
\centering
\includegraphics[width=3.5in]{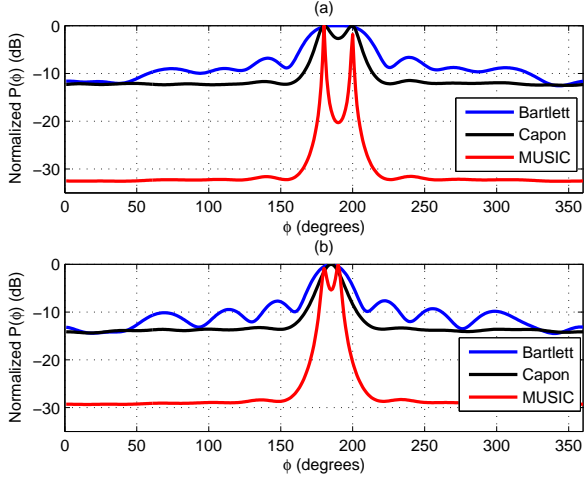}
\caption{Comparison of existing AoA estimation techniques for two sources with $M=16$, $N=1000$ samples and SNR = 0 dB: (a) $\phi_1=180$ and $\phi_2=200$  (b) $\phi_1=180$ and $\phi_2=190$ }
\label{fig_02}
\end{figure}

In Figure \ref{fig_02}, the resolution of the three techniques is investigated. We plot the normalized spatial power spectrum for two sources (a) with 20 degrees separation and (b) for 10 degrees separation. Again, the MUSIC technique outperforms the two other techniques achieving a resolution of almost 10 degrees with PFR = 5 dB where Capon achieved 20 degrees with PFR = 3 dB and Bartlett failed to resolve the two sources for a separation of less than 30 degrees. The superiority of the MUSIC comes on the cost that it requires an extensive computational complexity and that the number of sources must be known a priori or estimated. The estimation can be done using techniques such as Akaike information criterion (AIC) and minimum description length (MDL) \cite{Djuric96}, which adds extra computation complexity burden to the system. Also, MUSIC requires that the sources must be uncorrelated.

\section{Proposed Cross-Correlation Switched Beam system (XSBS)}

The existing AoA estimation techniques either have a low resolution problem or require extensive computational complexity to estimate the AoA. Moreover, they require $M$ receivers to implement the AoA estimation technique which increases the hardware complexity tremendously. Although the conventional SBSs have both low hardware and computational complexity, they fail to operate at medium and low level SNRs. This is mainly because the estimated AoA is the angle of the beam with the highest RSS.

We propose a novel cross-correlation based SBS (XSBS) AoA estimation technique. Our XSBS benefits from the low hardware complexity of the conventional SBS, which requires a single receiver, yet does not sacrifice the resolution or performance at medium and low level SNR. Moreover, our XSBS requires minimal computational complexity to estimate the AoA since it is based on estimating the cross correlation between two collected one dimensional vector of samples. With such low hardware and computational complexity, our XSBS will consume less power which will be very beneficial, particularly, if implemented on a portable device. Furthermore, our XSBS does not require neither any prior information on the number of the sources nor that the sources be uncorrelated.


\subsection{XSBS Design}
Our XSBS goes through two phases to estimate the AoA: \begin{itemize}
                                                                                                  \item \textbf{Phase I}: in this phase the \textit{Weights Unit} depicted in Figure \ref{fig_000} sends the first set of weights $w_o$. The set $w_o$ is applied on the antenna elements such that no phase shift or attenuation is applied at selected antenna elements, hence these selected antenna elements operate as omni-directional antennas or stand alone antennas. At the same time the weights are applied at the antenna elements in between the selected antenna elements to minimize their contribution to the summed signal to zero. The number of diversity antennas, $M_o$, as well as the separation between them depends on the antenna array formation. The objective is to include as many diversity antennas as possible, yet not collecting a correlated signals. The XSBS then acquires $N$ samples to collect the signal $x_o$.

                                                                                                  \item \textbf{Phase II}: In this phase the omni-directional signal collected in the first phase, i.e., $x_o$, becomes our reference signal.  The \textit{Weights Unit} sends the sets of weights $w_k$, for $k \in [1:K]$. The set $w_k$ steers the main beam of the antenna array to the direction $\phi_k$. The XSBS then acquires $N$ samples to collect the signal $x_k$.

                                                                                                      The next step is the core step in our XSBS. A cross correlation operation between our reference signal $x_o$ and the $k^{th}$ beam signal,$x_k$ is applied. The cross correlation operation is a simple operation as we will show later on. Therefore, the computational complexity of our XSBS is still low.                                                                                                 \end{itemize}

As we stated earlier, our XSBS is not exclusive to a particular antenna array formation. Rather, it can operate with any formation, linear, circular, planar or any antenna array formation where the steering vector of the antenna can be estimated either analytically or experimentally. An adequate separation between the antenna elements is needed to minimize the spatial correlation between the signal collected from each of them. Since the antenna elements are already placed $0.5\lambda$ apart, a limited number of antenna elements can be used as omni-directional diversity antennas. As $M_o$ increases, the XSBS can estimate signal with lower SNR values.

Additionally, the total number of antenna elements in the array is a key factor in determining the resolution of our XSBS. The higher the number of antenna elements, the smaller the half power beam width (HPBW) of the antenna array beam. A smaller HPBW leads to a better resolution. On the contrary, a higher number of antenna elements will increase the hardware complexity of the XSBS since they will require more components to apply the weights.

\begin{figure}
\centering
\includegraphics[width=3.5in]{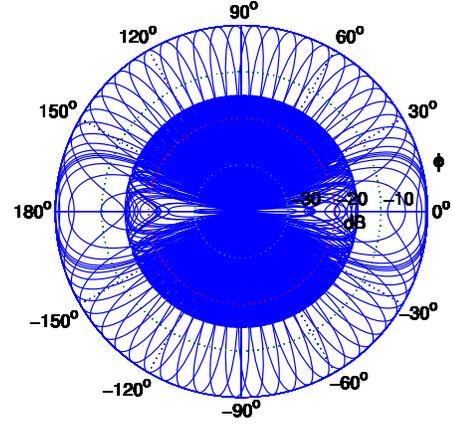}
\caption{Beam switching antenna array for $N= 17$ with Dolph-Chebychev excitation and $R= 15$ dB and $d=0.5\lambda$ with a total of 32 orthogonal beams with HBPW = 6 degrees}
\label{fig2}
\end{figure}

Using a non-uniform excitation such as Dolph-Chebyshev excitation, it is possible to generate more \textit{orthogonal} beams for the same $M$. The weights generated based on the Dolph-Chebyshev polynomials are given by \cite{AoA_book_Orfandis}:
\begin{align}
w_k(\phi)=W_{M-1}(y)
\end{align}
where $y=y_o\cos(0.5\phi)$. $y_o$ is estimated as:
\begin{align}
y_o=\cosh\left(\frac{acosh(R)}{M-1}\right)
\end{align}
where $R$ is the mainl obe to side lobe ratio.
\begin{align}
W_{M-1}(y)=\cos\left((M-1)acos(y)\right)
\end{align}
To study the resolution of our XSBS, we plot the steered antenna array beam for $M=17$, separation $d=0.5\lambda$, $R=15$ dB,  with Dolph-Chebyshev non-uniform excitation in Fig. \ref{fig2}. The achieved HPBW is approximately 6 degrees with a total of $K=32$ \textit{orthogonal} beams scanning the 180 degrees\footnote{Fig. \ref{fig2} is plotted using the MATLAB toolbox of \cite{AoA_book_Orfandis}.}. As $M$ increases, the resolution of the XSBS improves since the HPBW decreases. For a linear array or a planar array, as $M$ increases, $M_o$ increases. For our $M=17$ linear antenna array, $M_o = 5$ with a separation of $2\lambda$. For a circular array, the maximum achievable separation between the antenna elements is the diameter of the circle, which is likely $=\lambda$. Therefore, $M_o$ should not be $>2$.

\subsection{Cross Correlation Estimation}

After the completion of the first phase of our XSBS is completed by collecting an omni-directional signal, our XSBS moves to the second phase of estimating the AoA. In this phase, the omni-directional signal, $x_o$, collected earlier now becomes our XSBS reference signal. Our XSBS then starts to scan the angular region of interest and collect the the signals $x_k$, for $k\in[1:K]$:
\begin{align}
  x_k(n)=\sum_{M}{w_k s(n)}+v(n)
\end{align}

The \textit{Weights Unit} sends the precalculated weight sets, $w_k$, to the antenna array such that the main beam is directed towards the $k^{th}$ angle. The cross correlation coefficient between our reference signal and the $k^{th}$ signal can be given by:
\begin{align}
R_{ko}=\frac{1}{N}\sum_{n=1}^{N}{x_k(n)x_o^H(n)} \label{eqn11}
\end{align}
It is worth noting that Eq. (\ref{eqn11}) is applied on two vectors each has a dimension of $1 \times N$, while the autocovariance function in Eq. (\ref{eqn10}) is applied on a matrix with a dimension $M\times N$. For the MUSIC algorithm, an eigen-decomposition operation is then applied on this autocovariance function along with the other steps explained in the earlier section to estimate the AoA. For our XSBS, Eq. (\ref{eqn11}) is all the computation needed to estimate the AoA, which tremendously reduces the computational complexity.

\subsection{Performance Evaluation}

We evaluate the performance of our XSBS under different SNR levels, number of samples and for different antenna array formation. We compare the performance of our XSBS to the MUSIC algorithm, which we showed earlier to outperform the conventional techniques.

\begin{figure}
\centering
\includegraphics[width=3.5in]{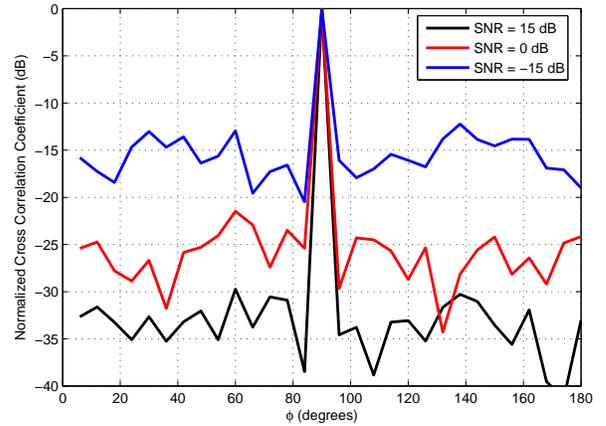}
\caption{Normalized cross correlation coefficient for different SNR for $M= 17$, $M_o=5$ and $N = 1000$ samples at different SNR levels.}
\label{fig3}
\end{figure}

In Fig. \ref{fig3}, we simulate our XSBS with linear antenna array with Dolph-Chebyshev excitation with $M=17$, $M_o=5$ and $N=1000$ for a signal arriving $\phi_k=90^\circ$ at different SNR levels. We plot the normalized cross correlation coefficient Eq. (\ref{eqn11}), which is now our spatial power, versus the azimuth angle $\phi$. It is shown that our XSBS has a superb performance achieving a PFR = 33 dB at SNR = 27 dB , PFR = 27 dB at SNR = 0 dB and PFR = 17 at SNR = -15 dB.

\begin{figure}
\centering
\includegraphics[width=3.5in]{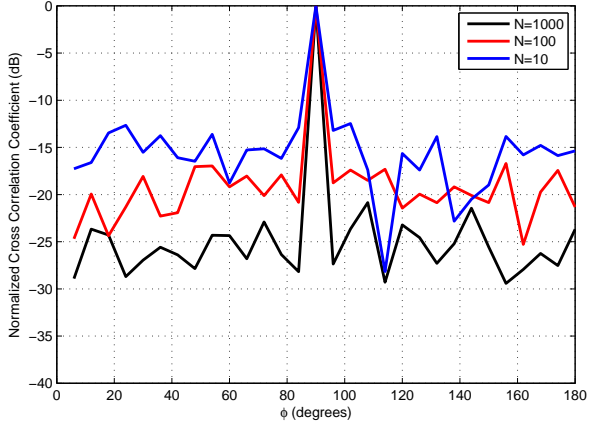}
\caption{Normalized cross correlation coefficient for different number of samples for SNR=0 dB for $M= 17$, $M_o=5$ for $N=10,100$ and $1000$ samples.}
\label{fig30}
\end{figure}

We then study the effect of changing the number of collected samples $N$ on the performance of our XSBS. In Fig. \ref{fig30} we simulate our XSBS with linear antenna array with Dolph-Chebshev excitation with $M=17$, $M_o=5$ and SNR = 0 dB for a signal arriving $\phi_k=90$ for $N=10,100$ and $1000$ samples. As expected, as the number of samples increases, the performance of our XSBS improves. From figures \ref{fig3} and \ref{fig30}, it also can be inferred that at higher SNR values, a lower number of samples is required to achieve an adequate PFR.

Ths MUISC algorithm requires that the number of sources be known a priori. The maximum number of sources for the MUSIC algorithm is limited by the number of antenna elements $M$. In other words the MUSIC algorithm can not estimate the AoA if the number of sources is $>M$. In addition to that, the MUSIC algorithm requires that the signals from different sources must be uncorrelated. Our XSBS is limited by the total number of beams $K$ which is much larger than $M$. Also, our XSBS does not require any prior information about the number of sources. Moreover, our XSBS does not require that the signals from different sources be uncorrelated.

\begin{figure}
\centering
\includegraphics[width=3.5in]{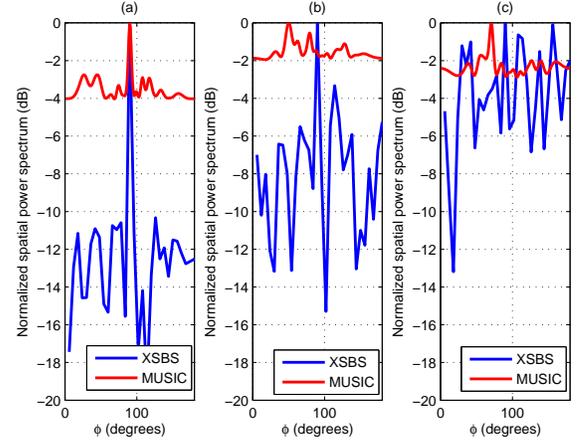}
\caption{Spatial power spectrum for our XSBS vs. the MUSIC algorithm at: (a) SNR=-20 dB, (b) SNR = -25 dB and (c) SNR = -30 dB}
\label{fig401}
\end{figure}

We compare the performance of our XSBS to the MUSIC algorithm in Fig. \ref{fig401} for the same number of samples of $N=1000$ at very low SNR levels: (a) SNR=-20 dB, (b) SNR = -25 dB and (c) SNR = -30 dB. The MUSIC algorithm in this figure used a uniform linear array with $M=16$, while our XSBS uses an $M=17$ linear array with Dolphy-Chebyshev excitation. It is shown that the MUSIC algorithm fails to estimate the AoA at SNR levels $<20$ dB with PFR = 3dB, while our XSBS can operate at SNR levels as low as -25 dB with PFR = 5 dB.

\begin{figure}
\centering
\includegraphics[width=3.5in]{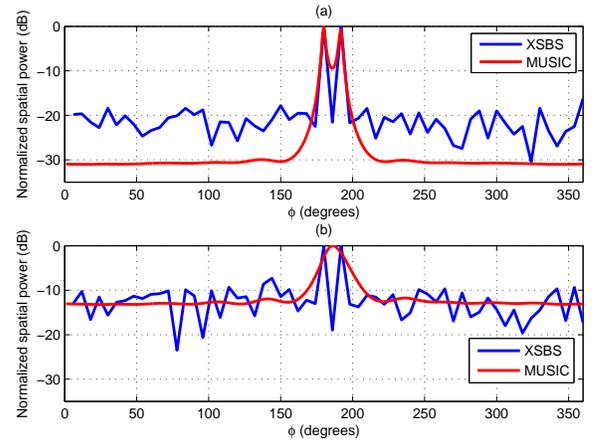}
\caption{Cross correlation coefficient for different number of samples for SNR=-20 dB for M= 17 and 5 omni directional beams}
\label{fig402}
\end{figure}

In Fig. \ref{fig402}, we compare the resolution of our XSBS to the resolution of the MUSIC algorithm. We use a uniform circular array for the two techniques. We plot the spatial power spectrum for two sources arriving at angles $\phi_1$ and $\phi_2$. The two sources for the MUSIC are uncorrelated while we use the same signal for the two sources for our XSBS. We plot the spatial power spectrum for two sources at $\phi_1=180^\circ$ and $\phi_2=192^\circ$ at: (a) SNR = 0 dB and (b) SNR = -15 dB. It is shown that the resolution of the MUSIC highly depends on the received SNR, while for our XSBS, it depends on the HPBW of the main lobe. The resolution of the MUSIC algorithm is about $8^\circ$ with PFR = 3dB at SNR = 0 dB and $20^\circ$ at SNR = -15 dB. Our XSBS has a consistent resolution of the HPBW, which in this case is $6^\circ$.
\section{Conclusion}
In this paper, we proposed a hardware friendly AoA estimation system. Our system first collects an omni-directional signal to be used as a reference signal. Our system then switches the main beam to scan the angular region of interest. The collected signals from the switched beams are cross correlated with the reference signal. The cross correlation coefficient is the highest at the true AoA and relatively negligible otherwise. Our algorithm can operate with any antenna array formation with known steering vector. We showed that our algorithm can operate at very low SNR level, i.e., as low as - 25 dB. As the SNR increases, a fewer samples can be used to achieve an adequate PFR. The number of sources that can be detected using our system is limited by the number of switched beams, which is higher than the number of antenna elements. We compared our algorithm to the MUSIC algorithm and we showed that our algorithm has a superior performance over it. Not only because our algorithm does not require any prior information or constraints on the received signal, but also it has a better resolution. The resolution of our system mainly depends on the HPBW of the main beam rather than the SNR such as MUSIC. Unlike the MUSIC algorithm, our system requires a single receiver, which reduces the hardware complexity tremendously. Moreover, our system is based on estimating the cross correlation coefficient between one dimensional vector, which has a negligible computational complexity when compared to estimating the eigenvalue decomposition of an autocovariance matrix.
\section*{Acknowledgment}
This research was made possible by NPRP 5-559-2-227 grant from the Qatar National Research Fund (a member of The Qatar Foundation). The statements made herein are solely the responsibility of the authors.

\bibliographystyle{IEEEtran}
\bibliography{references_AoA_estimation}
\end{document}